\documentclass{article}
\usepackage[utf8]{inputenc}
\usepackage{amssymb}
\usepackage{amsfonts}
\usepackage{amsmath}
\usepackage{bm}
\usepackage{mathtools}
\usepackage{commath}
\usepackage{array}
\usepackage{esvect}
\usepackage{cite}
\usepackage[dvipsnames]{xcolor}
\usepackage{physics}
\usepackage{mathrsfs}
\usepackage{authblk}
 \usepackage{multirow}
\usepackage{geometry}
\usepackage{abstract}
\usepackage[english]{babel}
\usepackage[T1]{fontenc}
\usepackage[colorlinks,urlcolor=blue,citecolor=blue]{hyperref}
 \newcommand{\sfrac}[2]{{\textstyle\frac{#1}{#2}}}

\usepackage[all]{xy}

\renewcommand{\=}{\ =\ }
\newcommand{\im}{\textrm{i}}
\newcommand{\ep}{\textrm{e}}

\DeclareMathAlphabet\mathbfcal{OMS}{cmsy}{b}{n}
\DeclareMathAlphabet{\boldmathe}{T1}{cmr}{bx}{it}
\newcommand{\mbf}[1]{\boldmathe{#1}}
\def\C{\mathbb C}
\def\R{\mathbb R}
\def\Z{\mathbb Z}
\def\vx{\mbf{x}}
\def\vz{\mbf{z}}

\def\vd{\bm{\partial}}

\begin{document}

\begin{center}
{\LARGE \bf
Coupling and particle number intertwiners \\ in the Calogero model \\
}
\vspace{6mm}
{\Large Francisco Correa$^a$, Luis Inzunza$^{a}$, Olaf Lechtenfeld$^{b}$
}
\\[6mm]
\noindent ${}^a${\em 
Departamento de F\'isica, Universidad de Santiago de Chile, \\ Av. Victor Jara 3493, Santiago, Chile}\\[3mm]
\noindent ${}^b${\em Institut f\"ur Theoretische Physik and Riemann Center for Geometry and Physics\\
Leibniz Universit\"at Hannover \\
Appelstra\ss{}e 2, 30167 Hannover, Germany} \\ [3mm]

{\tt francisco.correa.s@usach.cl, luis.inzunza@usach.cl, olaf.lechtenfeld@itp.uni-hannover.de}

\vspace{8mm}
\end{center}

\begin{abstract} 

\noindent
It is long known that quantum Calogero models feature intertwining operators, which increase or decrease the coupling constant
by an integer amount, for any fixed number of particles. We name these as ``horizontal'' and construct new ``vertical'' intertwiners,
which \emph{change the number of interacting particles} for a fixed but integer value of the coupling constant.
The emerging structure of a grid of intertwiners exists only in the algebraically integrable situation (integer coupling) and allows one 
to obtain each Liouville charge from the free power sum in the particle momenta by iterated intertwining either horizontally or vertically.
We present recursion formul\ae\ for the intertwiners as a factorization problem for partial differential operators
and prove their existence for small values of particle number and coupling.
As a byproduct, a new basis of non-symmetric Liouville integrals appears, algebraically related to the standard symmetric one.

\end{abstract}

\section{Introduction and summary}

The Calogero--Sutherland model (for short: Calogero model) was introduced in 1971 in its simplest form 
($A_{n-1}$ type rational or trignometric) as a fully solvable quantum system of $n$~interacting nonrelativistic identical particles 
on a line \cite{Cal69,suther}.  
This model has found applications in diverse fields, including black hole physics \cite{Gibbons:1998fa}, 
hydrodynamics \cite{Kulkarni:2017jal}, solitons \cite{OlshaPere81-rev}, 
among many others \cite{OlshaPere83-rev, Poly06-rev, bmodels, cmsbook}.

Calogero models became a paradigm of integrable systems, since they feature all the various properties characteristic for 
(in fact, maximal super)integrability \cite{Woj,Kuz}.  
What is more, two Calogero models with coupling strengths $\ell=\hbar^2g(g{-}1)$ and $\ell'=\hbar^2g'(g'{-}1)$ are isospectral 
(up to a few missing states) whenever $g{-}g'\in\Z$. 
This is implemented by so-called intertwining (or shift) operators~$M_n(g)$ of differential order~$\tfrac12n(n{-}1)$, 
which map the eigenstates of a Calogero Hamitonian~$H_n(g)$ to those of~$H_n(g{+}1)$ \cite{Opdam, ChaVes90, Heckman}. 
As a consequence, for integral values of~$g$ the energy spectrum of these models is explicitly computable by 
applying a finite sequence of intertwiners to the non-interacting (free) $n$-particle system (at $g{=}1$ or~$0$). 
This property holds in fact for all $n$ conserved Liouville charges, of which the total momentum~$P$ and energy~$H$ are only the first two.
(Maximal superintegrability implies $n{-}1$ additional conserved charges not in involution, which we do not consider here.)
Plotting from left to right the increasing sequence of $g=1,2,3,\ldots$ for a fixed number~$n$ of particles, 
we denote $M_n(g)$ as ``horizontal intertwiners''.

Exploiting the invariance of all Liouville integrals under the coupling flip $g\leftrightarrow1{-}g$, 
the sequence~$Q$ of $2g{-}1$ intertwiners leading from $H_n(1{-}g)$ to $H_n(g)$ for $g{\ge}1$ comes full circle 
and provides an \emph{additional} and algebraically independent conserved charge in involution \cite{ChaVes90, Chalykh96, vsc},
which is totally \emph{anti}symmetric under particle exchange.
Including $Q$ as a differential operator of order~$\tfrac12n(n{-}1)(2g{-}1)$, 
the complete ring of commuting differential operators corresponding to the Liouville charges 
gets enhanced to a ``supercomplete'' one, and one speaks of ``supersymmetric'' or ``algebraic'' integrability 
\cite{Opdam, ChaVes90, Heckman,Chalykh96,vsc,clp}. 

In this paper, we construct ``vertical intertwiners'' relating the spectra of \emph{same-coupling} Calogero Hamiltonians
with \emph{different particle number}. In other words, we provide strong evidence for operators~$W_n(g)$ 
of differential order~$n(g{-}1)$ which intertwine $H_n(g)$ (plus a free particle) with~$H_{n+1}(g)$. 
Falling short of a general proof, we give examples for $g{=}3$ at $n{=}2$ and~$3$.
We also provide an existence proof for~$W_2(g{\le}7)$.
Moreover, the vertical intertwiners function not only for the Hamiltonian but also for the other symmetric Liouville charges.
It is important to note, though, that the $W_n(g)$ only exist for integral values of~$g$, 
hence they are intimately tied to the algebraic integrability.
Altogether, all Calogero Hamiltonians $H_n(g)$ on an integral $(n,g)$ lattice are connected by intertwiners, 
and any one may be reached either horizontally from the free $H_n(1)$ system 
or vertically from the free $H_1(g)$ (plus $n{-}1$ free particles) system, or from any other free Hamiltonian
by different sequences of horizontal and/or vertical intertwiners.

As a byproduct, we find for $g\in\Z$ an unusual basis $\{P,\mathcal{S}_n^k\}$ with $k=1,\ldots,n$ 
of $n{+}1$ Liouville integrals for~$H_n(g)$, emphasizing again the algebraic integrability. 
While the $n$~standard Liouville integrals are totally symmetric under particle exchange,
i.e.~invariant under the Weyl/Coxeter reflection group~$S_n$ of~$A_{n-1}$,
the~$\mathcal{S}_n^k$ are invariant only under an~$S_{n-1}$ subgroup.
Their ring, however, includes polynomials in the totally symmetric charges as well as in the antisymmetric~$Q$.
We give several explicit examples and speculate on the relation between the two bases.

This paper is organized as follows. 
Section~\ref{sec2} provides a review of the key integrable aspects of the rational Calogero model, including the Liouville charges, 
horizontal intertwiners, and algebraic integrability. 
In Section~\ref{sec3}, we present the main results supporting the existence of vertical intertwiners for integer coupling. 
Section~\ref{sec4} introduces a novel basis for the Liouville integrals constructed from the vertical intertwiners. 
Finally,~Section \ref{sec5} offers a discussion and outlines several open problems.

\section{Horizontal intertwiners and algebraic integrability}\label{sec2}

The $n$-particle rational Calogero model lives on the phase space $\{x_i,p_i\}$ with $i=1,\ldots,n$ 
and is defined in terms of inverse-square interactions,
\begin{equation}\label{hcal}
H_{n}(g)\=\sum_{i=1}^n p_i^2\ +\ \sum_{i<j}^n\frac{2g(g{-}1)}{(x_i-x_j)^2} \ ,
\end{equation}
where we notice the $A_{n-1}$ Coxeter group and its root system as the underlying symmetry structure. 
Both standard and algebraic integrability at the quantum level can be established in terms of the so-called 
Dunkl or exchange operators \cite{dunkl,Poly1},
\begin{equation}
\pi_\ell(g)\=p_\ell\ +\ \im \sum_{j(\neq\ell)}^n \frac{g}{x_\ell-x_j} s_{\ell j}\ , \quad  [\pi_\ell, \pi_m]=0 \ ,
\end{equation}
where $p_\ell=-\im\partial_\ell$, and the symbols $s_{\ell m}$ denote the permutation operators satisfying
\begin{equation}
s_{\ell j} x_j=x_\ell s_{\ell j}\  , \quad s_{\ell j} p_j=p_\ell s_{\ell j} \ , \quad s_{\ell j}^2=1  \ .
\end{equation}
With the Dunkl operators one is able to construct the conserved charges and the standard intertwining operators. 
For example, the Liouville charges can be obtained as the power sums in the Dunkl operators,
\begin{equation}
I_r(n,g)\= {\rm res } \sum_{i=1}^n \pi_i^r(g)
\end{equation}
where the notation $ {\rm res }(\cdot) $ represents the restriction to totally symmetric functions. 
After pushing the permutation operators to the right, the symmetric restriction trivializes their action so that they can be neglected, 
creating the Liouville charges as purely local products of momenta and coordinates, i.e.~as differential operators of order~$r$.\footnote{
For simplicity of notation, in the following we sometimes suppress the dependence on the particle number~$n$ and/or the coupling~$g$ 
and indicate only the degree of the charge, $I_r(n,g)=I_r$.} 
The three lowest Liouville charges are
\begin{align}
I_1 &\= \sum_i p_i \ , \\
I_2 &\ \equiv\ H_n(g)\=\sum_{i=1}^n p_i^2\ +\ \sum_{i<j}^n\frac{2g(g{-}1)}{(x_i-x_j)^2} \ , \\
I_3 &\= \sum_{i=1}^n p_i^3\ +\ \sum_{i<j}^n\frac{3g(g{-}1)}{(x_i-x_j)^2}(p_i+p_j)\ .
\end{align}
It is possible to separate the free center-of-mass degree of freedom and remain with expressions 
depending only on the \emph{differences} of momenta and coordinates, effectively reducing to $n{-}1$ phase-space degrees of freedom. 
On the level of the Liouville integrals, this is achieved by a basis change,
\begin{equation}
I_r \ \to\ \ \bar{I}_r = I_r -\ \textrm{a polynomial in $\{I_{<r}\}$} \qquad\textrm{for}\ r\ge2
\end{equation}
so that $[\bar{I}_r,\sum_j x_j]=0$~\cite{cllm}.
The conformal symmetry and its corresponding algebra provides additional non-commuting conserved charges, 
rendering the system quantum superintegrable, see for example~\cite{Woj,Kuz,clp,ccl}. 
However, in this paper we will employ only the algebraic Liouville integrability. 

While the symmetric combinations of Dunkl operators yield the Liouville charges, with antisymmetric combinations of Dunkl operators 
we can build intertwining operators. The simplest such choice is a Vandermonde-type product,
\begin{equation}
M_n(g)\= {\rm res } \prod_{\ell<m} \bigl(\pi_\ell(g)-\pi_m(g)\bigr)  \ .
\end{equation}
This totally antisymmetric operator is of differential order $n(n{-}1)/2$ independent of~$g$, 
and it intertwines the Calogero Hamiltonian increasing the coupling constant in the following way,
\begin{equation}\label{hi1}
M_n(g)\,H_{n}(g) \= H_{n}(g{+}1)\,M_n(g)  \ ,
\end{equation}
which we represent in terms of the following \emph{horizontal action} diagram,
\begin{equation}\label{hori}
\xymatrixcolsep{4pc}
\xymatrixrowsep{3pc}
\xymatrix{
H_{n}(g) \ar[r]^{\!\!\!\!\!\!  M_{n}(g)}          &H_{n}(g{+}1)} \ .
\end{equation}
Any horizontal intertwiner is, in fact, an algebraic expression in~$M_n(g)$ and the~$I_r$.
The symmetry $g\leftrightarrow1{-}g$ of the Hamiltonian allows one to directly construct the inverse intertwiner,
\begin{equation}
M_n(g)^\dagger=M_n(-g) \ , \qquad M_n(g)^\dagger\,H_{n}(g{+}1) \= H_{n}(g)\,M_n(g)^\dagger \  ,
\end{equation}
which can be represented as
\begin{equation}\label{horil}
\xymatrixcolsep{4pc}
\xymatrixrowsep{3pc}
\xymatrix{
H_{n}(g)        &H_{n}(g{+}1)  \ar[l]^{  M_{n}(g)^\dagger}  } \ .
\end{equation}
The action of these intertwiners actually extends to all Liouville integrals \cite{clp},
\begin{equation}\label{hi2}
M_n(g)\,I_r(n,g) \= I_r(n,g{+}1)\,M_n(g)\ , \qquad M_n(g)^\dagger\,I_r(n,g{+}1) \= I_r(n,g)\,M_n(g)^\dagger  \ .
\end{equation}

In the spirit of supersymmetric quantum mechanics, where the product of two supercharges gives a Hamiltonian (or a polynomial therein), 
a successive application of both intertwinings in~(\ref{hi2}) must result in a totally symmetric conserved quantity,
\begin{equation}\label{hpol}
M_n(g)^\dagger M_n(g) = \mathcal{R}_n(g) \ , \qquad 
M_n(g)\,M_n(g) ^\dagger= \mathcal{R}_n(g{+}1)\ , \qquad
[\mathcal{R}_n(g), I_r(n,g)]=0 \ ,
\end{equation}
where the operators $\mathcal{R}_n(g)$ of differential order~$n(n{-}1)$ were explicitly given in~\cite{clp}.
Since the center of mass is absent in the intertwiners~$M_n(g)$, the $\mathcal{R}_n(g)$
are certain polynomials in the reduced Liouville integrals~$\bar{I}_r$, 
whose coefficients are independent of~$g$ since we may evaluate them in the limit of infinitely separated particles.
We may therefore write
\begin{equation}
\mathcal{R}_n(g) = R_n(\bar{I}_2,\bar{I}_3,\ldots,\bar{I}_n) \equiv R_n(\bar{I})\ .
\end{equation}
For illustration, we recall some explicit expressions, 
abbreviating $x_{ij}=x_i{-}x_j$ and $\partial_{ij}=\partial_i{-}\partial_j$:
\begin{equation}
\label{M1(g)}
\begin{aligned}
\im M_2(g)&\=\partial_{12}-\frac{2g}{x_{12}}\ ,\\
-\im M_3(g)&\=
\partial_{12}\partial_{23}\partial_{31}
-\frac{2g}{x_{31}}\partial_{12}\partial_{23}-\frac{2g}{x_{23}}\partial_{31}\partial_{12}-\frac{2g}{x_{12}}\partial_{23}\partial_{31} \\
&\quad+\Bigl( \frac{4 g^2 }{x_{31} x_{23}} -\frac{g (g{-}1)}{x_{12}^2} \Bigr)\partial_{12}
+\Bigl( \frac{4 g^2 }{x_{12} x_{31}}-\frac{g (g{-}1) }{x_{23}^2}\Bigr) \partial_{23}
+\Bigl( \frac{4 g^2 }{x_{12} x_{23}} -\frac{g (g{-}1)}{x_{31}^2} \Bigr) \partial_{31} \\
&\quad-\frac{6 g^2 (g{+}1)}{x_{12} x_{31} x_{23}}+2 g (g{-}1) (g{+}2)  \Bigl(\frac{1}{x_{12}^3}+\frac{1}{x_{31}^3}+\frac{1}{x_{23}^3} \Bigr)\ .
\end{aligned}
\end{equation}
The corresponding conserved polynomials read~\cite{clp}
\begin{equation} 
\label{Rn}
\begin{aligned}
R_2 &\= 2\bar{I}_2 \qquad\qquad\quad\!\textrm{with}\quad \bar{I}_2=I_2-\tfrac12I_1^2\ ,\\
R_3 &\= -3\bar{I}_3^2+\tfrac12\bar{I}_2^3\qquad\textrm{with}\quad
\bar{I}_2 = I_2-\tfrac13I_1^2\ ,\quad \bar{I}_3=I_3-I_2I_1+\tfrac29I_1^3\ ,\\
R_4 &\= -4\bar{I}_4^3+\bar{I}_4^2\bar{I}_2^2+2\bar{I}_4\bar{I}_3^2\bar{I}_2-\tfrac13\bar{I}_3^4-\tfrac49\bar{I}_3^2\bar{I}_2^3\\
& \qquad\textrm{with}\quad \bar{I}_2 = I_2-\tfrac14I_1^2\ ,\quad
\bar{I}_3 = I_3-\tfrac34I_2I_1+\tfrac18I_1^3\ ,\quad
\bar{I}_4 = I_4-I_3I_1-\tfrac14I_2^2+\tfrac12I_2I_1^2-\tfrac{1}{16}I_1^4\ .
\end{aligned}
\end{equation}

The horizontal intertwiners are also useful to reveal the algebraic integrability at integer values of the coupling~$g$. 
In this case, it is possible to find an extra Hermitian conserved charge $Q_n(g)$ 
thanks to the symmetry $g\leftrightarrow1{-}g$ of the Hamiltonian, 
\begin{equation}
Q_n(g)\=M_n(g{-}1)\,M_n(g{-}2)\cdots M_n(1)\,M_n(0)\,M_n(1)^\dagger\cdots M_n(g{-}2)^\dagger\,M_n(g{-}1)^\dagger\ , \qquad 
Q_n(g)=Q_n(g)^\dagger \ .
\end{equation}
$Q_n(g)$ is of differential order $\frac{1}{2}n(n{-}1)(2g{-}1)$ and commutes with all Liouville charges, $[Q_n(g),I_r(n,g)]=0$.
Being totally antisymmetric under particle exchange, it is algebraically independent of the~$I_r$,
thus extending the Liouville ring according to algebraic integrability.
Its square however is given in terms of the polynomials $\mathcal{R}_n$,
\begin{equation}
Q_n(g)^2\=\bigl(\mathcal{R}_n(g)\bigr)^{2g-1} \ ,
\end{equation}
while the horizontal intertwining relation for this ``odd'' charge reads
\begin{equation}
M_n(g)\,Q_n(g)\,\mathcal{R}_n(g) \= Q_n(g{+}1)\,M_n(g)\ .
\end{equation}
The algebraic properties of all conserved charges (including those not in involution) have been studied explicitly 
for the rational Calogero model at $n=3$ and $n=4$ in~\cite{clp}. 
The details for other Coxeter groups and trigonometric variants together with non-Hermitian deformations 
were explored in~\cite{ccl,Correa:2019hnu}, see also the review~\cite{alptrev}.

For later use, we need to add to the $n$-particle Calogero Hamiltonian~(\ref{hcal}) an $(n{+}1)$st particle without any interaction,
thereby trivially extending the Hamiltonian to
\begin{equation}\label{hcalp}
H_{n}^{+1}(g)\=H_{n}(g)+p_{n+1}^2 \ .
\end{equation}
Since $M_n(g)$ does not depend on this additional free particle, the intertwining relation (\ref{hi1}) is not affected, 
\begin{equation}
M_n(g)\, H_{n}^{+1}(g)\= H_{n}^{+1}(g{+}1)\,M_n(g) \ ,
\end{equation}
extending the diagram~(\ref{hori}) to 
\begin{equation}\label{hori2}
\xymatrixcolsep{4pc}
\xymatrixrowsep{3pc}
\xymatrix{
H_{n}^{+1}(g) \ar[r]^{\!\!\!\!\!\! M_{n}(g)}  & H_{n}^{+1}(g{+}1)} \ .
\end{equation}
Of course, the same is true for the reverse action.
Although this modified Hamiltonian looks innocuous, it will be relevant to what comes next.

\section{Vertical intertwiners and algebraic integrability}\label{sec3}

\subsection{A result of Chalykh, Feigin and Veselov}

The new findings we put forward in this paper were triggered by old results of Chalykh, Feigin and Veselov in~\cite{defo} 
(and studied later in \cite{kp, ceo, khodarinova}), revealing a one-parameter family of quantum integrable 
Calogero Hamiltonians~$\widetilde{H}_{n+1}(g)$ different from~$H_{n+1}(g)$ in~(\ref{hcal}). 
These models were generated from~$H_n^{+1}(g)$ by certain intertwining operators and are algebraically integrable for integer values of~$g$, 
but they have no classical counterpart. Their Hamiltonian has the form
\begin{equation}\label{hcfv1}
\begin{aligned}
\widetilde{H}_{n+1}(g)&\=
\sum_{i=1}^{n+1} p_i^2+\sum_{i<j}^n\frac{2\hbar^2g(g{-}1)}{(x_i-x_j)^2} +\sum_{i=1}^n \frac{2\hbar^2g}{(x_i-\sqrt{g{-}1}\, x_{n+1})^2} \\ 
&\= H_{n}^{+1}(g)+\sum_{i=1}^n \frac{2\hbar^2g}{(x_i-\sqrt{g{-}1}\, x_{n+1})^2}\ .
\end{aligned}
\end{equation}
We have temporarily installed $\hbar$ to illustrate that the $n{+}1$st particle with the non-Calogero interation decouples in the classical limit
$\hbar\to0$ with $\hbar g$ finite.
The key to the integrability of~(\ref{hcfv1}) is the existence of an intertwining operator $\widetilde{W}_n(g)$
of differential order $n(g{-}1)$, which connects 
$\widetilde{H}_{n+1}(g)$ to $H_n^{+1}(g)$ of~(\ref{hcalp}),
\begin{equation}\label{iwt}
\widetilde{W}_n(g)\,H_{n}^{+1}(g)\=\widetilde{H}_{n+1}(g)\,\widetilde{W}_n(g) \ .
\end{equation}
These models were extended to other Coxeter root systems and to trigonometric and elliptic interactions~\cite{defo}. 

Here, we are interested only in the particular case $g{=}2$, where $\sqrt{g{-}1}=1$ brings the $n{+}1$st-particle interaction 
back to Calogero form,
\begin{equation}
\widetilde{H}_{n+1}(2) \= \sum_{i=1}^{n+1} p_i^2+\sum_{i<j}^n\frac{4}{(x_i-x_j)^2} +\sum_{i=1}^n \frac{4}{(x_i-x_{n+1})^2} \= H_{n+1}(2)\ .
\end{equation}
Hence, for this value of the coupling, the intertwining
\begin{equation} \label{intn2}
W_n(2)\,H_{n}^{+1}(2)\=H_{n+1}(2)\,W_n(2) \qquad\textrm{for}\quad W_n(2) \equiv \widetilde{W}_n(2)
\end{equation}
increases the number of Calogero interactions from the Hamiltonian (\ref{hcalp}) to (\ref{hcal}) (with $n\to n{+}1$).
This suggests the existence of a more general structure underlying the standard Calogero models, 
in the form of \emph{vertical intertwiners} $W_n(g)\neq\widetilde{W}_n(g)$ for $g{>}2$, but also of differential order~$n(g{-}1)$. 

\subsection{Vertical intertwiners from two to three particles}

Let us validate our suggestion for the simplest case of $n{=}2$ particles, where
\begin{equation}\label{hcfv2}
\widetilde{H}_{3}(2)\=p_1^2+p_2^2+p_3^2 + \frac{4}{(x_1-x_2)^2} + \frac{4}{(x_1-x_3)^2} + \frac{4}{(x_2-x_3)^2} \= H_{3}(2)\ ,
\end{equation}
and
\begin{equation}\label{iwt2}
W_2(2)\,H_{2}^{+1}(2)\=H_{3}(2)\,W_2(2) \ .
\end{equation}
The explicit form of the known intertwiner reads~\cite{defo}
\begin{equation}\label{w22}
W_2(2)\=\partial_{13}\partial_{23}-\frac{2}{x_{13}}\partial_{23}-\frac{2}{x_{23}}\partial_{13}+ 
\Bigl(\frac{4}{x_{13} x_{23}}+\frac{2}{x_{12}^2}\Bigr)\ .
\end{equation}
As in the horizontal case, the reverse action is effected by the adjoint operator $W_2(2)^\dagger$,
namely
\begin{equation}\label{w22i}
W_2(2)^\dagger\=\partial_{13}\partial_{23}+\frac{2}{x_{13}}\partial_{23}+\frac{2}{x_{23}}\partial_{13}+ 
\Bigl(\frac{4}{x_{13} x_{23}}+\frac{2}{x_{12}^2}-\frac{2}{x_{13}^2}-\frac{2}{x_{23}^2}\Bigr)\ ,
\end{equation}
so that 
\begin{equation}
W_2(2)^\dagger\,H_{3}(2)\=H_{2}^{+1}(2)\,W_2(2)^\dagger \ ,
\end{equation}
We illustrate these actions as vertical intertwiners with the following diagrams, 
\begin{equation}\label{vert}
\xymatrixcolsep{4pc}
\xymatrixrowsep{3pc}
\xymatrix{
H_{2}^{+1}(2) \ar[d]^{W_2(2)} \\
H_{3}(2) } \qquad  \qquad \xymatrix{
H_{2}^{+1}(2) \\
H_{3}(2) \ar[u]^{W_2(2)^\dagger}  } \  .
\end{equation}

Combining the diagrams (\ref{hori}), (\ref{hori2}) and (\ref{vert}) we obtain the following scheme,
\begin{equation}
\xymatrixcolsep{4pc}
\xymatrixrowsep{3pc}
\xymatrix{
H_{2}^{+1}(2) \ar[d]^{W_2(2)} \ar[r]^{\!\!\! M_2(2)} & H_{2}^{+1}(3) \ar[d]^{W_2(3)\,?} \ar[r]^{\!\!\! M_2(3)} & H_{2}^{+1}(4) \ar[d]^{W_2(4)\,?} \ar[r]^{\ M_2(4)} & \cdots \\
H_{3}(2) \ar[r]^{M_{3}(2)} & H_{3}(3) \ar[r]^{M_{3}(3)} & H_{3}(4) \ar[r]^{\ M_{3}(4)} & \cdots  } 
\end{equation}
The question marks indicate the natural proposition of operators which generalize (\ref{vert}) 
to higher integral values of the coupling $g$,\footnote{
Recall that the intertwiners $\widetilde{W}_2(g)$ of~\cite{defo} for $g{>}2$ produce non-Calogero interactions for the additional particle.}
\begin{equation} \label{int2g}
W_2(g)\,H_{2}^{+1}(g)\=H_{3}(g)\,W_2(g)  \qquad\textrm{and}\qquad
W_2(g)^\dagger\,H_3(g)\=H_2^{+1}(g)\,W_2(g)^\dagger\ .
\end{equation}
We will argue that such operators do exist and can be obtained recursively by solving a certain factorization problem.
The commutativity of the above diagram implies that 
\begin{equation} \label{fact}
M_{3}(g)\,W_2(g)\=W_2(g{+}1)\,M_2(g) \qquad\textrm{for}\quad g=2,3,4,\ldots
\end{equation}
or, for reversed arrow directions,
\begin{equation}
W_2(g)^\dagger\,M_3(g)^\dagger\=M_2(g)^\dagger\,W_2(g{+}1)^\dagger\ .
\end{equation}
Naively, one might want to solve (\ref{fact}) simply by writing
\begin{equation}
W_2(g{+}1) \ \buildrel{?}\over{=}M_{3}(g)\,W_2(g)\,M_2(g)^\dagger \qquad\Rightarrow\qquad
W_2(g{+}1)\,M_2(g) \= M_{3}(g)\,W_2(g)\,\mathcal{R}_2(g)
\end{equation}
but this yields an operator of differential order $2(g{+}1)$ rather than~$2g$, 
which does not quite reproduce the left-hand side of~(\ref{fact}).
Therefore, given $W_2(g)$, (\ref{fact}) constitutes a factorization problem for~$W_2(g{+}1)$, 
which we claim has a unique solution.

While we do not have a closed formula or recursion for the vertical intertwiners~$W_2(g)$,
we can prove our proposition up to~$g{=}7$ (see the Appendix \ref{app1}). 
Since the center of mass is not affected by the vertical interwtining~(\ref{int2g}),
these operators are composed of~$\{p_{ij},x_{\ell m}\}$.
We have explicitly constructed them up to~$g{=}5$.
Explicitly, for $g{=}3$ and $g{=}4$,
\begin{equation}
\begin{aligned}
W_2(3) &\= \partial_{13}^2 \partial_{23}^2
-\frac{6}{x_{13}} \partial_{13} \partial_{23}^2
-\frac{6}{x_{23}} \partial_{13}^2 \partial_{23}
+\frac{12}{x_{13}^2} \partial_{23}^2
+\frac{12}{x_{23}^2} \partial_{13}^2
+\Bigl(\frac{36}{x_{13} x_{23}}+\frac{12}{x_{12}^2}\Bigr)\partial_{13} \partial_{23} 
\\[4pt] &\quad
-\Bigl(\frac{72}{x_{13}^2 x_{23}}+\frac{36}{x_{12}^2 x_{13}}+\frac{12}{x_{12}^3}\Bigr)\partial_{23}
-\Bigl(\frac{72}{x_{23}^2 x_{13}}+\frac{36}{x_{12}^2 x_{23}}-\frac{12}{x_{12}^3}\Bigr)\partial_{13}
+\frac{144}{x_{12}^2}\Bigl(\frac{1}{x_{13}^2}+\frac{1}{x_{23}^2}-\frac{1}{x_{13}x_{23}}\Bigr)\ ,
\end{aligned}
\end{equation}
and

\begin{equation} \label{W24}
\begin{aligned}
W_{2}(4) &\=
\frac{1}{2}\partial _{13}^3 \partial _{23}^3-
\frac{12}{x_{13}}\partial _{13}^2\partial _{23}^3 +
\frac{60 }{x_{13}^2}\partial _{13} \partial _{23}^3-\frac{120}{x_{13}^3}\partial _{23}^3
+ 18 \Bigl(\frac{4}{x_{13} x_{23}}+\frac{1}{x_{12}^2}\Bigr)\partial _{13}^2 \partial _{23}^2
\\[4pt]&\quad
+\alpha_1(x)\partial_{13}\partial_{23}^2  
+\alpha_2(x)\partial _{13}^2
+\alpha_3(x)\partial _{13} \partial _{23}
+\alpha_4(x)\partial _{23}
+\alpha_5(x)+\text{cyclic(1,2)}\ ,
\end{aligned}
\end{equation} 
where coefficients $\alpha_i(c)$ are given in the Appendix \ref{app2}.  

\subsection{Vertical intertwiners beyond three particles}

Of course, this structure is not restricted to including a third particle into the Calogero interactions. 
Going from three to four particles in quite similar to the previous case but a little more involved since the differential order
increases from $2(g{-}1)$ to $3(g{-}1)$.
The simplest case $g{=}2$ can be read off from~\cite{defo},
\begin{equation}
\begin{aligned}
W_{3}(2) &\= \partial_{14}\partial_{24}\partial_{34}
-\frac{2}{x_{34}} \partial_{14}\partial_{24}-\frac{2}{x_{24}}\partial_{14}\partial_{34}
-\frac{2}{x_{14}} \partial_{24}\partial_{34}
\\ & \quad
+\Bigl(\frac{4}{x_{24} x_{34}}+\frac{2}{x_{23}^2}\Bigr)\partial_{14}
+\Bigl(\frac{4}{x_{14} x_{34}}+\frac{2}{x_{13}^2}\Bigr)\partial_{24}
+\Bigl(\frac{4}{x_{14} x_{24}}+\frac{2}{x_{12}^2}\Bigr)\partial_{34}
\\ & \quad
-4\Bigl(\frac{1}{x_{34} x_{12}^2}+\frac{1}{x_{24}x_{31}^2}+\frac{1}{x_{14} x_{23}^2}+\frac{2}{x_{14} x_{24}x_{34}}\Bigr)\ ,
\end{aligned}
\end{equation}
while for $g{=}3$ we obtained an additional intertwiner,
\begin{equation} \label{W33}
\begin{aligned}
W_3(3) &\= \frac{1}{3}\partial_{14}^2\partial_{24}^2 \partial_{34}^2-\frac{6}{x_{34}}\partial_{14}^2\partial_{24}^2\partial_{34}
+\frac{12}{x_{34}^2}\partial_{14}^2\partial_{24}^2+\Bigl(\frac{36}{x_{24} x_{34}}+\frac{12}{x_{23}^2}\Bigr)\partial_{14}^2\partial_{24}\partial_{34}
\\ & \quad
-\Bigl(\frac{36}{x_{23}^2 x_{24}}+\frac{72}{x_{24}^2 x_{34}}+\frac{12}{x_{23}^3}\Bigr)\partial_{14}^2\partial_{34}
+\frac{1}{3}\Bigl( \frac{72}{x_{24} x_{31}^2}+\frac{72}{x_{12}^2 x_{34}}+\frac{216}{x_{14} x_{24} x_{34}}+\frac{72}{x_{14} x_{23}^2}\Bigr)\partial_{14}\partial_{24}\partial_{34}
\\ & \quad
+q_3(x)\,\partial_{14}^2 +q_2(x)\,\partial_{14}\partial_{34}+q_1(x)\,\partial_{14}+\frac{1}{3}q_0(x)\ +\ \text{cyclic}(1,2,3)
\end{aligned}
\end{equation}
where the rational functions $q_0$, $q_1$, $q_2$ and $q_3$ are given in the Appendix \ref{app2}.

It is now clear how the general picture for any number~$n$ of particles looks like.
As our main result, we conjecture for integer values of $g$ the existence of vertical intertwining operators~$W_n(g)$ 
and their adjoints~$W_n(g)^\dagger$ satisfying
\begin{equation}\label{iwtn}
W_n(g)\,H_{n}^{+1}(g)\={H}_{n+1}(g)\,W_n(g) \qquad\textrm{and}\qquad
W_n(g)^\dagger\,H_{n+1}(g)\=H_n^{+1}(g)\,W_n(g)^\dagger\ ,
\end{equation}
respectively, completing the scheme in a generic way,
\begin{equation}\label{generic}
\xymatrixcolsep{4pc}
\xymatrixrowsep{3pc}
\xymatrix{
\cdots \ar[r]^{\!\!\!\!\!\!\! M_n(g{-}1)} &
H_{n}^{+1}(g) \ar[d]^{W_n(g)} \ar[r]^{\!\!\!\!\!\! M_n(g)} & 
H_{n}^{+1}(g{+}1) \ar[d]^{W_n(g{+}1)} \ar[r]^{\quad M_n(g{+}1)} & \cdots \\
\cdots \ar[r]^{\!\!\!\!\!\! M_{n+1}(g{-}1)} &
H_{n+1}(g) \ar[r]^{\!\!\!\!\! M_{n{+}1}(g)} & 
H_{n+1}(g{+}1) \ar[r]^{\quad\ M_{n{+}1}(g{+}1)} & \cdots  } 
\end{equation}
What is more, the intertwining~(\ref{iwtn}) extends to all standard Liouville charges,
\begin{equation}\label{wconi}
W_n(g)\,I_r^{+1}(n,g)\=I_r(n{+}1,g)\,W_n(g)  
\qquad \text{where}\quad  I_r^{+1}(n,g)\=p_{n+1}^r+I_{r}(n,g) \quad\textrm{for}\  r=1,\ldots, n{+}1\  ,
\end{equation}
analogous to the horizontal action given in equation (\ref{hi2}).
For $r=n{+}1$ the charge $I_{n+1}(n,g)$ is of course algebraically dependent in the $n$-particle system,
but the additional~$p_{n+1}$ accounts for the correct number $n{+}1$ of commuting conserved charges.
Obviously, the $I_r^{+1}(n,g)$ are totally symmetric in particle indices ranging only from $1$ to~$n$.
In Appendix~\ref{app3} we present a proof of~(\ref{wconi}) based on (\ref{iwtn}) and~(\ref{hi2}), 
employing some ideas from Appendix~\ref{app1}.
Less clear is a possible intertwining relation of $W_n(g)$ with the odd charge~$Q_n(g)$.

The scheme implies that $W_n(g{+}1)$ is a differential operator of order $n\,g$ and can be found from~$W_n(g)$ by solving the relation
\begin{equation} \label{factorize}
M_{n+1}(g)\,W_n(g)\=W_n(g{+}1)\,M_n(g)\ .
\end{equation}
In other words, given the left-hand side one should split off a right factor of $M_n(g)$ to obtain the vertical intertwiner at coupling~$g{+}1$.
The corresponding relation for the adjoint intertwiner is
\begin{equation}
W_n(g)^\dagger\,M_{n+1}(g)^\dagger\=M_n(g)^\dagger\,W_n(g{+}1)^\dagger\ .
\end{equation}
As already emphasized, all intertwiners are translation invariant.
This structure is intimately connected with the algebraic integrability of the Calogero model.
It remains an open problem to establish a constructive existence proof for any integer value of $n$ and~$g$.

\subsection{Calogero models from free particles via vertical intertwinings}

The generic scheme~(\ref{generic}) can be extended vertically by adding more than a single free particle at a time.
In this way, we may start with $n{+}1$ non-interacting particles and intertwine vertically all the way to $H_{n+1}(g)$.
Also, the diagram ends on the left with the coupling $g{=}1$, corresponding again to $n{+}1$ free particles.
In this way, we end up with the extended diagram
\begin{equation}\label{extended}
\xymatrixcolsep{4pc}
\xymatrixrowsep{3pc}
\xymatrix{
H_1^{+n}(1) \ar@{=}[d] \ar@{=}[r] & H_1^{+n}(2) \ar[d]^{W_1(2)} \ar@{=}[r] & H_1^{+n}(3) \ar[d]^{W_1(3)} \ar@{=}[r] & H_1^{+n}(4) \ar[d]^{W_1(4)} \ar@{=}[r] & \ldots \\
H_2^{+(n-1)}(1) \ar@{=}[d] \ar[r]^{M_2(1)} & H_2^{+(n-1)}(2) \ar[d]^{W_2(2)} \ar[r]^{M_2(2)} & H_2^{+(n-1)}(3) \ar[d]^{W_2(3)} \ar[r]^{M_2(3)} & H_2^{+(n-1)}(4) \ar[d]^{W_2(4)} \ar[r]^{\quad M_2(4)} & \ldots \\
H_3^{+(n-2)}(1) \ar@{=}[d] \ar[r]^{M_3(1)} & H_3^{+(n-2)}(2) \ar[d]^{W_3(2)} \ar[r]^{M_3(2)} & H_3^{+(n-2)}(3) \ar[d]^{W_3(3)} \ar[r]^{M_3(3)} & H_3^{+(n-2)}(4) \ar[d]^{W_3(4)} \ar[r]^{\quad M_3(4)} & \ldots \\
\vdots \ar@{=}[d] & \vdots \ar[d]^{W_{n-1}(2)} & \vdots \ar[d]^{W_{n-1}(3)} & \vdots \ar[d]^{W_{n-1}(4)} & \\
H_n^{+1}(1) \ar@{=}[d] \ar[r]^{M_n(1)} & H_n^{+1}(2) \ar[d]^{W_n(2)} \ar[r]^{M_n(2)} & H_n^{+1}(3) \ar[d]^{W_n(3)} \ar[r]^{M_n(3)} & H_n^{+1}(4) \ar[d]^{W_n(4)} \ar[r]^{\quad M_n(4)} & \ldots \\
H_{n+1}(1) \ar[r]^{M_{n+1}(1)} & H_{n+1}(2) \ar[r]^{M_{n+1}(2)} & H_{n+1}(3) \ar[r]^{M_{n+1}(3)} & H_{n+1}(4) \ar[r]^{\quad M_{n+1}(4)} & \ldots }
\end{equation}
where the outer double lines are only identifications. By choosing an appropriate path in this diagram,
one may relate two Calogero models with particle numbers $n_1\le n_2$ and integer couplings $g_1$ and $g_2$ via a chain of intertwiners leading from
$H_{n_1}^{+(n_2-n_1)}(g_1)$ to $H_{n_2}(g_2)$, in many different ways.
In particular, a given Hamiltonian~$H_n(g)$ can be reached from the non-interacting $H_n(1)$ not only by iterating the horizontal intertwiners~$M_n$
but likewise by starting from the same free Hamiltonian~$H_1^{+(n-1)}(g)$ by successively applying the vertical intertwiners~$W(g)$.
It is interesting to note that iterated horizontal intertwining increases stepwise the Calogero interaction strength for all particles simultaneously,
while interated vertical intertwining turns on the final Calogero interaction for only one particle at a time.
This should open up new ways of computation for the rational Calogero models at integer couplings.

\section{Non-symmetric Liouville charges}\label{sec4}

In Section~2 we recalled that combining a horizontal intertwiner with its adjoint produced a particular homogeneous polynomial
in the Liouville charges, see~(\ref{hpol}). 
A natural question is whether a similar property holds for the vertical intertwiners and their adjoints.
Indeed, from the scheme~(\ref{generic}) one should expect to find for integral values of~$g$ a conserved charge 
$\mathcal{S}_{n+1}(g)$ of differential order $2n(g{-}1)$ for $H_{n+1}(g)$ via
\begin{equation}\label{hpol1}
W_n(g)\,W_n(g)^\dagger\= \mathcal{S}_{n+1}(g) \ , \qquad 
[\mathcal{S}_{n+1}(g),{H}_{n+1}(g)]=0 \ , \qquad 
[\mathcal{S}_{n+1}(g),I_r(n{+}1,g)]=0\ .
\end{equation}
The other composition, $W_n(g)^\dagger\,W_n(g)$, yields a conserved charge for~$H_n^{+1}(g)$, which seems less interesting.
More remarkable, however, is the fact that the charge~$\mathcal{S}_{n+1}(g)$ \emph{cannot} 
be expressed in terms of the Liouville charges~$I_r$ of the system, even though it is in involution with them. 

In order to understand this feature, we note an important point regarding the permutation symmetry of the Hamiltonians and intertwiners.
Let us come back to the $(n{=}2,g{=}2)$ case for a moment.
While $H_3(2)$ (and also $M_3(2)$) is totally symmetric in all three particle labels, $H_2^{+1}(2)$ (and also $M_2(2)$) 
singles out particle~3 and is only $1\leftrightarrow 2$ symmetric, and thus the same is true for $W_2(2)$ and $W_2(2)^\dagger$.
Selecting particle~3 as the one whose Calogero interaction is turned on via the action of~$W_2(2)$ is of course just 
one of three possible choices. Therefore, one has in fact three different vertical intertwiners ($k=1,2,3$)
\begin{equation}\label{severla}
W_2^{k}(2) \, : \qquad \text{Calogero interactions in}\,\, (x_i,x_j) \longrightarrow (x_i,x_j,x_k) 
\qquad\textrm{for $i$, $j$, $k$ all distinct} \ . 
\end{equation}
These are obtained by applying the `broken' permutations,
\begin{equation} \label{W123}
W_2^3(2) \equiv W_2(2)\ ,\qquad
W_2^2(2) = {\rm res} \left[ s_{23} W_2(2)\right]\ ,\qquad
W_2^1(2) = {\rm res} \left[ s_{13} W_2(2)\right]\ ,
\end{equation}
but of course link $H_3(2)$ to different manifestations of $H_2^{+1}(2)$.
For each of the three values of~$k$ we have a different conserved charge for $H_3(2)$,
\begin{equation}
\begin{aligned}
\mathcal{S}^{1}_{3} (2)&\=W^{1}_2(2)W^{1}_2(2)^\dagger\ \propto\ \partial_{31}^2\partial_{12}^2\ +\ \text{lower terms} \ , \\%[4pt]
\mathcal{S}^{2}_{3} (2)&\=W^{2}_2(2)W^{2}_2(2)^\dagger\ \propto\ \partial_{12}^2\partial_{23}^2\ +\ \text{lower terms} \ ,\\%[4pt]
\mathcal{S}^{3}_{3} (2)&\=W^{3}_2(2)W^{3}_2(2)^\dagger\ \propto\ \partial_{23}^2\partial_{31}^2\ +\ \text{lower terms} \ . 
\end{aligned}
\end{equation}
These operators commute with one another, $[\mathcal{S}_3^k(2),\mathcal{S}_3^{k'}\!(2)]=0$, and thus function as Liouville charges,
It is explicit that they are non-symmetric and therefore cannot be expressed in terms of the standard reduced Liouville integrals 
$\bar{I}_2$ and~$\bar{I}_3$, which are totally symmetric by construction. 
Such a relation can only occur for symmetric combinations of the~$\mathcal{S}^k_3(2)$, for example
\begin{equation}
\begin{aligned}
\mathcal{S}^{1}_{3}(2)+\mathcal{S}^{2}_{3}(2)+\mathcal{S}^{3}_{3}(2) &\= \tfrac94\,\bar{I}_2^2 \ , \\[4pt]
\mathcal{S}^{1}_{3}(2)\,\mathcal{S}^{2}_{3}(2)+\mathcal{S}^{2}_{3}(2)\,\mathcal{S}^{3}_{3}(2)+\mathcal{S}^{3}_{3}(2)\,\mathcal{S}^{1}_{3}(2) 
&\= 3\,\bar{I}_2\,R_3(\bar{I}) \ , \\[4pt]
\mathcal{S}^{1}_{3}(2)\,\mathcal{S}^{2}_{3}(2)\,\mathcal{S}^{3}_{3}(2) &\= R_3(\bar{I})^2\ ,  \\[4pt]
\bigl(\mathcal{S}^{1}_{3}(2)\,\mathcal{S}^{2}_{3}(2)\bigr)^2+\bigl(\mathcal{S}^{2}_{3}(2)\,\mathcal{S}^{3}_{3}(2)\bigr)^2+\bigl(\mathcal{S}^{3}_{3}(2)\,\mathcal{S}^{1}_{3}(2)\bigr)^2 
&\= \tfrac92\,\bar{I}_2^2\,R_3(\bar{I})^2\ , 
\end{aligned}
\end{equation}
all in terms of the two polynomials
\begin{equation}
\bar{I}_2\equiv\bar{I}_2(3,2) \qquad\textrm{and}\qquad
R_3(\bar{I}) =- 3\,\bar{I}_3(3,2)^2+\tfrac12\,\bar{I}_2(3,2)^3
\end{equation}
given in (\ref{Rn}). Due to translation invariance only the reduced Liouville charges can appear.
However, it is also possible to form totally antisymmetric combinations, the simplest being
\begin{equation} \label{Santi}
\bigl(\mathcal{S}^{1}_{3}(2)-\mathcal{S}^{2}_{3}(2)\bigr)\,\bigl(\mathcal{S}^{2}_{3}(2)-\mathcal{S}^{3}_{3}(2)\bigr)\,\bigl(\mathcal{S}^{3}_{3}(2)-\mathcal{S}^{1}_{3}(2)\bigr) 
\= -9\,Q_3(2)\,\bar{I}_3
\end{equation}
with the antisymmetric extra Liouville charge~$Q$ present because of algebraic integrability for integral values of~$g$~\cite{clp}.

Generalizing to $(n{+}1{=}3,g{>}2)$ we have evaluated two kinds of symmetric polynomials in terms of $\mathcal{S}^{\ell}_{3} (g)$, namely
\begin{equation}
\begin{aligned}
\mathcal{S}^{1}_{3}(g)\,\mathcal{S}^{2}_{3}(g)\,\mathcal{S}^{3}_{3}(g) &\= R_3(\bar{I})^{2(g-1)} \ , \\[2pt]
\mathcal{S}^{1}_{3}(g)+\mathcal{S}^{2}_{3}(g)+\mathcal{S}^{3}_{3}(g)
&\ = \sum_{\ell=0}^{\lfloor\frac23(g-1)\rfloor}
 (-\tfrac32)^{2(g-1)-3\ell} \frac{2(g{-}1)\,\Gamma(2g{-}2{-}2\ell)}{\Gamma(\ell{+}1)\,\Gamma(2g{-}1{-}3\ell)}\,
\bar{I}_2^{2(g-1)-3\ell}\,R_3(\bar{I})^\ell \ .
\end{aligned}
\end{equation}
For the simplest asymmetric combination we have
\begin{equation}
\label{SQI}
\bigl(\mathcal{S}^{1}_{3}(g)-\mathcal{S}^{2}_{3}(g)\bigr)\,
\bigl(\mathcal{S}^{2}_{3}(g)-\mathcal{S}^{3}_{3}(g)\bigr)\,
\bigl(\mathcal{S}^{3}_{3}(g)-\mathcal{S}^{1}_{3}(g)\bigr) 
\= -9\,Q _3(g)\,U_{3}({\bar{I},g})\ ,
\end{equation}
where the first polynomials $U_{3}({\bar{I},g})$ are
\begin{equation}
\begin{aligned}
U_3(\bar{I},2) &\= \bar{I}_3 \ ,  \\%[3pt]
U_3(\bar{I},3) &\=3 \bar{I}_3^3+\sfrac{25 }{4} \bar{I}_3 \bar{I}_2^3\ ,\\%[3pt]
U_3(\bar{I},4) & \= 81\bar{I}_3^3 \bar{I}_2^3 +\sfrac{1323}{64}\bar{I}_3\bar{I}_2^6 \ , \\
U_3(\bar{I},5) &\=-27 \bar{I}_3^7+ \sfrac{783}{4} \bar{I}_3^5 \bar{I}_2^3+\sfrac{17667}{32}  \bar{I}_3^3  \bar{I}_2^6+\sfrac{7225}{128}\bar{I}_3  \bar{I}_2^9\ . \\
\end{aligned}
\end{equation}

Stepping the particle number up to $n{+}1=4$, we encounter four intertwiners~$W_3^k(g)$ with $k=1,\ldots,4$ and hence four non-symmetric
Liouville charges~$\mathcal{S}_4^k(g)$. For their analogous symmetric polynomials we find
\begin{equation}
\begin{aligned}
\mathcal{S}^{1}_{4}(g)\,\mathcal{S}^{2}_{4}(g)\,\mathcal{S}^{3}_{4}(g)\,\mathcal{S}^{4}_{4}(g) &\= R_4(\bar{I})^{2(g-1)} \ , \\[2pt]
\mathcal{S}^{1}_{4}(g)+\mathcal{S}^{2}_{4}(g)+\mathcal{S}^{3}_{4}(g)+\mathcal{S}^{4}_{4}(g) &\= 4(-1)^{g+1}T_4(\bar{I},g)\ ,
\end{aligned}
\end{equation}
with $R_4$ written in~(\ref{Rn}) and
\begin{equation}
\begin{aligned}
T_4(\bar{I},2) &\= \bar{I}_4 \bar{I}_2 + \sfrac{7}{9} \bar{I}_3^2 \ ,  \\%[3pt]
T_4(\bar{I},3) &\= 4\bar{I}_4^3 + \bar{I}_4^2 \bar{I}_2^2 + \tfrac{74}{9} \bar{I}_4 \bar{I}_3^2 \bar{I}_2 + \tfrac{61}{81} \bar{I}_3^4 + \tfrac49 \bar{I}_3^2 \bar{I}_2^3 \ ,\\%[3pt]
T_4(\bar{I},4) &\= 12\bar{I}_4^4 \bar{I}_2 + \tfrac{92}{3}\bar{I}_4^{3} \bar{I}_3^2 + \bar{I}_4^3 \bar{I}_2^3 + \tfrac{115}{3} \bar{I}_4^2 \bar{I}_3^2 \bar{I}_2^2
+ \tfrac{577}{27} \bar{I}_4 \bar{I}_3^4 \bar{I}_2 + \tfrac43 \bar{I}_4 \bar{I}_3^2 \bar{I}_2^4 +\tfrac{547}{729} \bar{I}_3^6 + \tfrac{92}{27} \bar{I}_3^4 \bar{I}_2^3
\ ,
\end{aligned}
\end{equation}
and so on. 

Clearly, for any given particle number~$n{+}1$ the Calogero system $H_{n+1}(g)$ is reached in $n{+}1$ ways 
by turning on the Calogero interaction for the $n{+}1$st particle in~$H_{n}^{+1}(g)$ 
via the vertical intertwiners~$W_n^k(g)$ for $k=1,\ldots,n{+}1$. 
Therefore, this system features $n{+}1$ non-symmetric Liouville charges
\begin{equation}
\mathcal{S}^{k}_{n+1}(g)\=W^{k}_n(g)\,W^{k}_n(g)^\dagger \ , \qquad  
[\mathcal{S}^{k}_{n+1}(g),H_{n+1}(g)]=0 \ , \qquad  
[\mathcal{S}^{k}_{n+1}(g),\mathcal{S}^{k'}_{n+1}(g)]=0 \ .
\end{equation}
Their number surpasses by one the number~$n$ of reduced symmetric Liouville charges~$\{\bar{I}_2,\bar{I}_3,\ldots,\bar{I}_{n+1}\}$.
However, the~$\mathcal{S}_{n+1}^k$ still form a \emph{complete non-symmetric basis} of Liouville integrals where, 
in contrast to the standard symmetric case, all integrals have the same differential order $2n(g{-}1)$. 
This is of course valid in the algebraically integrable situation of integer~$g$, 
where one has an extra \emph{antisymmetric} Liouville charge~$Q$, 
so that the full set $\{\bar{I}_2,\bar{I}_3,\ldots,\bar{I}_{n+1},Q\}$ is not totally symmetric anyway. 
The commutativity
\begin{equation}
[\mathcal{S}^{k}_{n+1}(g),\bar{I}_r(n{+}1,g)]=0 \quad\textrm{for}\ n=2,\ldots,n{+}1
\qquad\textrm{and}\qquad [\mathcal{S}^{k}_{n+1}(g), Q(g)]=0
\end{equation}
suggests that the set $\{\mathcal{S}_{n+1}^1,\mathcal{S}_{n+1}^2,\ldots,\mathcal{S}_{n+1}^{n+1}\}$ 
is related to the former by an algebraic transformation.\footnote{
We have verified this for the case of $n{+}1=3$ particles.}

\section{Discussion and outlook}\label{sec5}

The existence of the  vertical intertwiners opens up a whole new network of algebraic relations within the Calogero model.
Besides the concrete open points raised already in previous sections, 
let us widen the scope and mention some obvious further-reaching questions and possible generalizations.

The action of the horizontal intertwiner~$M_n(g)$ on the simultaneous eigenstates of the standard symmetric Liouville charges is well known.
In particular, in the presence of a confining harmonic potential these eigenspaces are finite-dimensional, 
and $M_n(g)$ maps them isospectrally, with a finite-dimensional kernel. 
It will be very interesting to analyze the action of the \emph{vertical} intertwiner~$W_n(g)$ on these eigenspaces and reveal its kernel.

The horizontal intertwiners can be used to formally extend the sequence of algebraically integrable models ``to the left'',
allowing for negative integer values of~$g$. While the $g\leftrightarrow1{-}g$ symmetry of the Liouville charges identifies
any such model with a corresponding one at $g\,{\ge}\,\tfrac12$, the energy eigenstates do not share this symmetry,
and so its application leads to an approximate doubling of eigenstates.  The additional states, however, are unphysical
because even in a confining harmonic potential they are not normalizable.
The vertical intertwiners~$W_n(g)$, in contrast, cannot even formally be extended to $g{<}0$, since their differential order~$n(g{-}1)$
would become negative. Still, their action on the unphysical states for $g{>}1$ can be studied.

For a more complete picture, the behavior of the non-involutive conserved charges under the vertical intertwiner
should be investigated. Since these charges are tied to the conformal invariance of the rational Calogero model,
we expect the conformal sl(2,$\R$) algebra to play a role here.

The $n$ non-symmetric Liouville charges $\mathcal{S}_n^k$ in the $n$-particle Calogero system with the center of mass removed
all have the same (even) differential order~$2(n{-}1)(g{-}1)$, depending on the value of~$g$. 
This is to be contrasted with the ($g$-independent) orders $r=2,\ldots,n$ of the reduced symmetric Liouville charges~$\bar{I}_r$ 
and the order $\tfrac12n(n{-}1)(2g{-}1)$ of the antisymmetric charge~$Q$.
Therefore, only polynomials of sufficiently high degree in the~$\bar{I}_r$ can be generated from the~$\mathcal{S}_n^k$,
but never the elementary charges~$\bar{I}_r$ themselves. Likewise, totally antisymmetric combinations such as~(\ref{Santi})
cannot produce~$Q$ without some ``dressing'' by a polynomial in~$\bar{I}_r$ .
Therefore, the totally (anti)symmetric Liouville ring and the new non-symmetric one only partially overlap in a subring.
A precise characterization of this subring must be left to future investigation.

Of course, the concept of a vertical intertwiner might apply to more general algebraically integrable models,
such as the deformations discussed in~\cite{sphere1, TA, sphere2, TA2, Correa:2023uwj},
rational Calogero models based on other Coxeter reflection groups,
or even trigonometric, hyperbolic or elliptic Calogero systems. 
It is also interesting to explore the existence of vertical intertwiners in the context of angular reductions~\cite{HNY, FeLePo13, CoLe15, CoLe17}. 
While horizontal intertwiners can be naturally constructed using Dunkl operators, we question whether a similar construction exists 
for vertical intertwiners. In particular, this leads to the investigation of Dunkl-type symplectic algebras~\cite{th}.

On a final note, we observe that the vertical intertwiner~$W_n$ relates two Calogero modes based on different Coxeter groups:
For
\begin{equation}
H_n^{+1} \= p_{n+1}^2 \ +\ \tfrac1n(p_1{+}p_2{+}\ldots{+}p_n)^2\ +\ \bar{H}_n
\end{equation}
one has the Weyl group of~$A_1\oplus A_{n-1}$, while intertwining with $W_n$ yields the Hamiltonian
\begin{equation}
H_{n+1} \= \tfrac{1}{n+1}(p_1{+}p_2{+}\ldots{+}p_n{+}p_{n+1})^2\ +\ \bar{H}_{n+1}
\end{equation}
carrying the Weyl group of~$A_1\oplus A_n$.
The $A_1$ parts refer to (different) center-of-mass components (written out explicitly),
but the crucial difference is the additional connected node in the Dynkin/Coxeter diagram of~$A_n$ versus~$A_{n-1}$,
due to the interactions of the $n{+}1$st particle being turned on.
It appears as if the vertical intertwiner adds a new node here.
It is amusing to speculate that such ``node-adding'' operators might exist more generally.

\subsection*{Acknowledgements}
F.C. and L.I. were supported by Fondecyt Grants No. 1211356 and No. 3220327, respectively. 
The authors acknowledge the warm hospitality and support of the Instituto de Ciencias F\'isicas y Matem\'aticas
at Universidad Austral de Chile in Valdivia, when this work was started.

\subsection*{Note added}
After submitting this article to the arXiv, Oleg Chalykh kindly informed us about \cite{BerestChalykh} 
where more general results have been obtained by different methods. 
In particular, the existence of our vertical intertwiners has been proven there as a special case.

%\newpage

\appendix

\section{Appendices}

\subsection{Existence test for $W_2(g)$}  \label{app1}

To establish the existence of the vertical intertwiners for any integral value of $n$ and~$g$
we must show the recursion~(\ref{factorize})
\begin{equation} \label{factor}
M_{n+1}(g)\,W_n(g) \= W_n(g{+}1)\,M_n(g)\ ,
\end{equation}
i.e.~prove that the left-hand side can be factored as claimed by the right-hand side.
Introducing the cumulated horizontal intertwiner
\begin{equation}
\mathbb{M}_n(g)\=M_n(g)\ldots M_n(2)\,M_n(1) 
\qquad\Rightarrow\qquad
\mathbb{M}_n(g)\,H_n(1) \= H_n(g{+}1)\,\mathbb{M}_n(g)
\end{equation}
one can iterate (\ref{factor}) and obtain
\begin{equation}
\begin{aligned}
W_n(g{+}1)\,\mathbb{M}_n(g) &\= W_n(g{+}1)\,M_n(g)\,M_n(g{-}1)\,M_n(g{-}2)\cdots M_n(1) \\
&\= M_{n+1}(g)\,W_n(g)\,M_n(g{-}1)\,M_n(g{-}2)\cdots M_n(1) \\
&\=M_{n+1}(g)\,M_{n+1}(g{-}1)\,W_n(g{-}1)\,M_n(g{-}2)\cdots M_n(1) \\
&\quad\vdots\\
&\=M_{n+1}(g)\,M_{n+1}(g{-}1)\,M_{n+1}(g{-}2)\cdots M_{n+1}(1)\,W_n(1)\\
&\= \mathbb{M}_{n+1}(g)
\end{aligned}
\end{equation}
since $W_n(1)=1$. Therefore, solving~(\ref{factor}) is equivalent to factorizing
\begin{equation} \label{Factor}
\mathbb{M}_{n+1}(g) \= W_n(g{+}1)\,\mathbb{M}_n(g)\ .
\end{equation}

While this appears to be quite involved for general~$n$, we try to prove it 
for $n{=}2$ but arbitrary integral $g$, 
\begin{equation} \label{Factor2}
M_3(g)\,W_2(g)\ = W_2(g{+}1)\,M_2(g) \qquad\Leftrightarrow\qquad
\mathbb{M}_3(g) \= W_2(g{+}1)\,\mathbb{M}_2(g)\ .
\end{equation}
Fortunately, the following mathematical theorem provides an existence test.

\noindent
\textbf{Theorem} \cite{Kasman}: 
Let $\vx=(x_1,\ldots,x_n)\in\C^n$ and $\vd=(\partial_1,\ldots,\partial_n)$, 
and consider partial differential operators
\begin{equation}
L=\sum_i r_i(\vx)\,p_i(\vd) \qquad\textrm{and}\qquad
M=\phi(\vx)\,q(\vd)\,\phi(\vx)^{-1}
\end{equation}
with (complex-valued) rational functions~$r_i$ and polynomials~$p_i$ making up~$L$,
while some differentiable function~$\phi$ and some irreducible polynomial~$q$ define~$M$.
Then, a factorization $L=W\,M$ with an unknown differential operator~$W$
holds if and only if
\begin{equation}
L\;\phi(\vx)\,\ep^{\vz\cdot\vx} \= 0 \qquad \forall\, \vz\in\C^n \quad\textrm{with}\quad q(\vz)=0
\end{equation}
where $\vz\cdot\vx$ is the usual scalar product in~$\C^n$.
In other words, for every~$\vz$ in the irreducible zero set of~$q$, 
the polynomial-exponential $\phi(\vx)\,\ep^{\vz\cdot\vx}$ must lie 
not only in the kernel of~$M$ (which is obvious) but also in the kernel of~$L$.

We shall apply this Theorem for~$n{=}3$ in order to recursively prove the existence of~$W_2(g)$.
To this end, we abbreviate $x=x_{12}$ and $\partial=\partial_{12}$ and introduce the monomials
\begin{equation}
\varphi_{l}(\vx)\= x^l \qquad\textrm{so that}\qquad M_2(g)\=\partial-\tfrac{2g}{x} \= \varphi_g\,\partial\,\varphi_g^{-1}
\end{equation}
is of the form required by the Theorem, with $\phi=\varphi_g$ and $q(\vz)=z_1{-}z_2$.
Therefore, the zero locus of~$q$ is parametrized by~$\vz=(k,k,0)$ with $k\in\C$.
It directly follows that
\begin{equation}
M_2(g)\,\varphi_l = 2(l{-}g)\,\varphi_{l-1} \qquad\Rightarrow\qquad
\mathbb{M}_2(g)\,\varphi_{2j-1} \= 2^{2g}(j{-}1)(j{-}2)\cdots(j{-}g)\,\varphi_{2j-1-g} \ .
\end{equation}
Hence, $\{\varphi_1,\varphi_3,\varphi_5,\ldots,\varphi_{2g-1}\}\in\mathrm{ker}\,\mathbb{M}_2(g)$.

Let us set up an induction in~$g$ starting with the case of~$g{=}1$.
It can be verified through direct computation that $\mathbb{M}_3(1)=M_3(1)$ annihilates 
$\varphi_1(\vx)\,\ep^{k\,x}=x\,\ep^{k\,x}$ for any~$k$. 
Therefore, the Theorem implies that
\begin{equation}
\mathbb{M}_3(1) \= W\;\mathbb{M}_2(1) \ ,
\end{equation}
and we may identify the unknown factor $W$ with $W_2(2)$. 

For the induction step let un now assume that the factorization~(\ref{Factor2}) holds for an arbitrary value of $g$, 
and try to factorize the next higher intertwiner,
\begin{equation}
\label{Wtilde1}
L\ \equiv\ \mathbb{M}_3(g{+}1) \= M_3(g{+}1)\,\mathbb{M}_3(g)
\= M_3(g{+}1)\,W_2(g{+}1)\,\mathbb{M}_2(g)\ =:\ \widetilde{L}\;\mathbb{M}_2(g)
\ \buildrel{?}\over{=}\ W\;\mathbb{M}_2(g{+}1)\ ,
\end{equation}
where the last equality is to be shown.
We see that the kernel of~$\mathbb{M}_2(g)$ is included in the kernel of~$L$. However, to establish the factorization
in the final (questioned) equality with the help of the Theorem we must consider the kernel of~$\mathbb{M}_2(g{+}1)$, 
which contains the additional function~$\varphi_{2g+1}$. To apply the Theorem once more, we thus must show that
\begin{equation} \label{test}
\mathbb{M}_3(g{+}1)\,\varphi_{2g+1}\,\ep^{k\,x} \ \buildrel{!}\over{=}\ 0 \qquad\Leftrightarrow\qquad
\widetilde{L}\;\mathbb{M}_2(g)\,\varphi_{2g+1}\,\ep^{k\,x} 
\= 2^{2g} g!\,\widetilde{L}\,\varphi_{g+1}\,\ep^{k\,x} \ \buildrel{!}\over{=}\ 0\ .
\end{equation}
Alternatively, because $\varphi_{g+1}\in\mathrm{ker}\,M_2(g{+}1)$, 
if $\varphi_{g+1}\,\ep^{k\,x}$ is also annihilated by~$\widetilde{L}$ 
then the Theorem demands that
\begin{equation}
\widetilde{L} \= W\,M_2(g{+}1) \qquad\Rightarrow\qquad L \= W\;\mathbb{M}_2(g{+}1)\ .
\end{equation}
Either way, we can identify $W=W_2(g{+}2)$. The recursion~(\ref{factor}) for $g\rightarrow g{+}1$ directly follows.

In general, verifying the equation (\ref{test}) for arbitrary $g$ can be challenging.  
Therefore, we have approached it as an existence test. In other words, 
we may determine whether the relation~(\ref{test}) holds by performing explicit computations for a given $g$. 
We have already checked this for $g$-values up to seven.
To extend this test to $n{>}2$, one has to express $M_n(g)$ in the form of the Theorem, which is not obvious.

\subsection{Rational functions in $W_2(4)$ and  $W_3(3)$}  \label{app2}

The coefficients of the lower terms for~$W_2(4)$ in (\ref{W24}) are
\begin{align}
\alpha_1(x) &\=\frac{72}{x_{12}}\Bigl(\frac{6}{x_{12} x_{13}}+\frac{10}{x_{13}^2}-\frac{10}{x_{12} x_{23}}-\frac{1}{x_{12}^2}\Bigr)\ , \\
\alpha_2(x)&\=\frac{72}{x_{12}}\Bigl(\frac{16}{x_{12}^2 x_{23}}-\frac{20}{x_{12}^2 x_{13}}-\frac{10}{x_{12} x_{23}^2}+\frac{20}{x_{23}^3}+\frac{1}{x_{12}^3}\Bigr)\ ,\\
\alpha_3(x)&\=\frac{36}{x_{12}^2}\Bigl(\frac{60}{x_{12} x_{13}}+\frac{50}{x_{13}^2}-\frac{60}{x_{12} x_{23}}+\frac{50}{x_{23}^2}+\frac{1}{x_{12}^2}\Bigr)\ ,\\
\alpha_4(x)&\=\frac{288}{x_{12}^2}\Bigl(\frac{45}{x_{12}^2 x_{23}}
-\frac{48}{x_{12}^2 x_{13}}-\frac{25}{x_{12} x_{13}^2}-\frac{25}{x_{13}^3}-\frac{25}{x_{12} x_{23}^2}-\frac{1}{x_{12}^3}
\Bigr)\ ,\\
\alpha_5(x)&\=\frac{144}{ x_{12}^3}\Bigl( \frac{168}{x_{12}^2 x_{23}}
-\frac{168}{x_{12}^2 x_{13}}-\frac{75}{x_{12} x_{13}^2}-\frac{75}{x_{12} x_{23}^2}-\frac{50}{x_{13}^3}+\frac{50}{x_{23}^3}-\frac{2}{x_{12}^3}
\Bigr)\ .
\end{align}

The rational functions appearing in (\ref{W33}) for $W_3(3)$ read
\begin{align}
q_3(x)&\=\frac{144}{x_{23}^2}\Bigl(
\frac{1}{x_{24}^2}+\frac{1}{x_{23} x_{24}}
-\frac{1}{x_{23} x_{34}}+\frac{1}{x_{34}^2}\Bigr)\ ,
\\ \nonumber
q_2(x)&\=72\left\{\frac{3}{x_{31}}\Bigl(\frac{1}{x_{23}^2 x_{24}}-\frac{1}{x_{12}^2 x_{24}}\Bigr)+\frac{1}{x_{14}}\Bigl(\frac{6}{x_{12}^2 x_{34}}+\frac{1}{x_{23}^3}\Bigr)+\frac{1}{x_{23}} \Bigl(\frac{6}{x_{24}^2 x_{31}}-\frac{3}{x_{12}^2 x_{34}}+\frac{1}{x_{12}^2 x_{23}}\Bigr)\right.
\\  &
\left. \qquad\qquad+\,
\frac{3 }{x_{12}}\Bigl(\frac{2}{x_{24}^2 x_{31}}+\frac{2}{x_{23}^2 x_{34}}-\frac{1}{x_{14} x_{23}^2}\Bigr)+\frac{2}{x_{24}^2 x_{31}^2}-\frac{1}{x_{12}^3 x_{34}}\right\}\ ,
\\ \nonumber
q_1(x)&\=72\left\{\frac{1}{x_{31}^2}\Bigl(\frac{1}{x_{12}^3 x_{31}}+\frac{2}{x_{24}^2 x_{31}}\Bigr)-
\frac{1}{x_{23}^2}\Bigl(\frac{1}{x_{12}^3}-\frac{2}{x_{14}^2 x_{23}}\Bigr)+
\frac{6 }{x_{12}}\Bigl(\frac{1}{x_{14}^2 x_{23}^2}-\frac{1}{x_{24}^2 x_{31}^2}\Bigr)\right.
\\ & \qquad\qquad
+\,\frac{3}{x_{12}^2} \biggl[\frac{\text{1}}{x_{14}}\Bigl(\frac{2}{ x_{31}^2}+\frac{1}{ x_{23}^2}\Bigr)+\frac{1}{x_{24}}\Bigl(\frac{1}{ x_{31}^2}+\frac{2}{ x_{23}^2}\Bigr)\biggr]
\\ &\left.\quad\qquad\quad
+\,\frac{12}{x_{12}^2} \biggl[\frac{\text{1}}{x_{23}}\Bigl(\frac{1}{ x_{24}^2}+\frac{1}{x_{12} x_{34}}-\frac{1}{x_{12} x_{24}}\Bigr)+\frac{\text{1}}{x_{31}}
\Bigl(\frac{1}{x_{12} x_{34}}-\frac{1}{ x_{14}^2}-\frac{1}{x_{12} x_{14}}\Bigr)
-\frac{\text{1}}{x_{34}}\Bigl(\frac{1}{ x_{31}^2}+\frac{1}{ x_{23}^2}\Bigr)\biggr]\right\} \ ,\nonumber
\\ \nonumber
q_0(x)&\=\frac{864}{x_{12} x_{23} x_{31} } \left\{
\frac{1}{x_{12}}\biggl[\frac{1}{x_{24}}\Bigl(\frac{1}{x_{31}}-\frac{2}{x_{24}}\Bigr)-\frac{1}{x_{14}}\Bigl(\frac{1}{x_{23}}+\frac{2}{x_{14}}\Bigr)+\frac{1}{x_{34}}\Bigl(\frac{1}{x_{23}}-\frac{1}{x_{31}}-\frac{2}{x_{34}}\Bigr)\biggr]
\right.\\&\qquad\qquad
+\,\frac{2}{x_{14}}\Bigl(\frac{1}{x_{31}^2}-\frac{1}{x_{14} x_{31}}-\frac{1}{x_{12}^2}\Bigr)
-\frac{2}{x_{31} x_{34}}\Bigl(\frac{1}{x_{34}}+\frac{1}{x_{31}}\Bigr)+\frac{2}{x_{24}}\Bigl(\frac{1}{x_{12}^2}-\frac{1}{x_{24} x_{31}}\Bigr)
\\&\nonumber\left.\qquad\qquad
+\,\frac{1}{x_{23}}\biggl[
\frac{1}{x_{14}}\Bigl(\frac{1}{x_{31}}-\frac{2}{x_{14}}\Bigr)-\frac{1}{x_{24}}\Bigl(\frac{1}{x_{31}}+\frac{2}{x_{24}}\Bigr)+\frac{2 }{x_{34}}\Bigl(\frac{1}{x_{23}}-\frac{1}{x_{34}}\Bigr)-\frac{2}{x_{23} x_{24}}\biggr]\right\}\ .
\end{align}

\subsection{Intertwining relations for $W_n(g)$ and the Liouville charges } \label{app3}
\label{AppInterI}
By combining relation (\ref{hi2}) together with (\ref{Factor})(which followed from (\ref{iwtn}) via~(\ref{factorize})) 
we construct the following chain of equations for any value of~$r$,
\begin{equation}
\begin{aligned}
W_n(g)\,I_r^{+1}(n,g)\,\mathbb{M}_n(g{-}1) &\=
W_n(g)\,\mathbb{M}_n(g{-}1)\,I_r^{+1}(n,1) \ \equiv\
W_n(g)\,\mathbb{M}_n(g{-}1)\,I_r(n{+}1,1) \\[4pt] \=
\mathbb{M}_{n+1}(g{-}1)\,I_r(n{+}1,1) &\=
I_r(n{+}1,g)\,\mathbb{M}_{n+1}(g{-}1) \=
I_r(n{+}1,g)\,W_n(g)\,\mathbb{M}_n(g{-}1)\ ,
\end{aligned}
\end{equation}
which implies
\begin{equation}
W_n(g)\,I_r^{+1}(n,g) \= I_r(n{+}1,g)\,W_n(g)\ .
\end{equation}
We remind that
\begin{equation}
I_r^{+1}(n,g)\=p_{n+1}^r+I_{r}(n,g)
\qquad\textrm{and}\qquad
I_{r}(n,g)\={\rm res } \sum_{j=1}^n \pi_j^{r}(g)\ .
\end{equation}
 For instance, in the case of $n{=}2$ we have 
\begin{align}
 I_1^{+1}(2,g)&\=p_1+p_2+p_3 \=  I_1(3,g)\ ,  \notag\\[2pt]
I_2^{+1}(2,g)&\=p_1^2+p_2^2+p_3^2 +\tfrac{2g(g{-}1)}{x_{12}^2} \ ,\qquad\qquad\quad\ \;
I_2(3,g) \= I_2^{+1}(2,g) + \tfrac{2g(g{-}1)}{x_{13}^2} + \tfrac{2g(g{-}1)}{x_{23}^2} \ ,\\ \notag
I_3^{+1}(2,g)&\=p_1^3+p_2^3+p_3^3+ \tfrac{3g(g{-}1)}{x_{12}^2}(p_1{+}p_2)\ ,\qquad
I_3(3,g) \= I_3^{+1}(2,g) + \tfrac{3g(g{-}1)}{x_{13}^2}(p_1{+}p_3) + \tfrac{3g(g{-}1)}{x_{23}^2}(p_2{+}p_3)\ .
\end{align}

\end{document}